# A Novel Mechanism for Detection of Distributed Denial of Service Attacks

Jaydip Sen

Innovation Lab, Tata Consultancy Services Ltd.
Bengal Intelligent Park, Salt Lake Electronic Complex, Kolkata 700091, India
Jaydip.Sen@tcs.com

**Abstract.** The increasing popularity of web-based applications has led to several critical services being provided over the Internet. This has made it imperative to monitor the network traffic so as to prevent malicious attackers from depleting the resources of the network and denying services to legitimate users. This paper has presented a mechanism for protecting a web-server against a distributed denial of service (DDoS) attack. Incoming traffic to the server is continuously monitored and any abnormal rise in the inbound traffic is immediately detected. The detection algorithm is based on a statistical analysis of the inbound traffic on the server and a robust hypothesis testing framework. While the detection process is on, the sessions from the legitimate sources are not disrupted and the load on the server is restored to the normal level by blocking the traffic from the attacking sources. To cater to different scenarios, the detection algorithm has various modules with varying level of computational and memory overheads for their execution. While the approximate modules are fast in detection and involve less overhead, they have lower detection accuracy. The accurate modules involve complex detection logic and hence involve more overhead for their execution, but they have very high detection accuracy. Simulations carried out on the proposed mechanism have produced results that demonstrate effectiveness of the scheme.

## 1 Introduction

A *denial of service* (DoS) attack is defined as an explicit attempt by a malicious user to consume the resources of a server or a network, thereby preventing legitimate users from availing the services provided by the system. The most common DoS attacks typically involve flooding with a huge volume of traffic and consuming network resources such as bandwidth, buffer space at the routers, CPU time and recovery cycles of the target server. Some of the common DoS attacks are SYN flooding, UDP flooding, DNS-based flooding, ICMP directed broadcast, Ping flood attack, IP fragmentation, and CGI attacks [1]. Based on the number of attacking machines deployed to implement the attack, DoS attacks are classified into two broad categories: (i) a single intruder consumes all the available bandwidth by generating a large number of packets operating from a single machine, or (ii) the distributed case where multiple attackers coordinate together to produce the same effect from several machines on the network. The latter is referred to as DDoS attack and owing to its distributed nature, it is very difficult to detect.

In this paper, a robust mechanism is proposed to protect a web server from DDoS attack utilizing some easily accessible information in the server. This is done in such a way that it is not possible for an attacker to disable the server host and as soon as the overload on the server disappears, the normal service quality resumes automatically. The detection algorithm has several modules that provide flexibility in deployment. While the approximate detection modules are based on simple statistical analysis of the network traffic and involve very less computational and memory overhead on the server, the accurate detection module is based on a statistical theory of hypothesis testing that has more overhead in its execution.

The rest of the paper is organized as follows: Section 2 presents some existing work in the literature on defense against DoS attacks. Section 3 describes the components of the proposed security system and the algorithms for detection and prevention of attacks. Section 4 presents the simulation results and the sensitivity analysis of the parameters of the algorithms. Section 5 concludes the paper while highlighting some future scope of work.

## 2  Related Work

Protection against DoS attacks highly depends on the model of the network and the type of attack. Several mechanisms have been proposed to solve the problem of DoS attacks. Most of them have weaknesses and fail under certain circumstances.

*Network ingress filtering* is a mechanism proposed to prevent attacks that use spoofed source addresses [2]. This involves configuring the routers to drop packets that have illegitimate source IP addresses. *ICMP traceback* messages are useful to identify the path taken by packets through the Internet [3]. This requires a router to use a very low probability with which traceback messages are sent along with the traffic. Hence, with sufficiently large number of messages, it is possible to determine the route taken by the traffic during an attack. *IP traceback* proposes a reliable way to perform hop by hop tracing of a packet to the attacking source from where it originated [4]. Yaar et al. have proposed an approach, called *path identifier* (Pi), in which a path fingerprint is embedded in each packet, enabling a victim to identify packets traversing the same paths through the Internet on a per packet basis, regardless of source IP address spoofing [5]. *Pushback* approaches have been proposed to extract attack signatures by rate-limiting the suspicious traffic destined to a congested link [6][7]. Mirkovic et al. have proposed a scheme named D-WARD that performs statistical traffic profiling at the edge of the networks to detect new types of DDoS attacks [8]. Zou et al. have presented an adaptive defense system that adjusts its configurations according to the network conditions and attack severity in order to minimize the combined cost introduced by false positives [9]. *Client side puzzle* and other *pricing algorithms* are effective tools to make protocols less vulnerable to depletion attacks of processing power [10].

## 3 The Major System Components and Algorithms

This Section describes the traffic model and the attack model on which the proposed security system has been designed. One of the most vital components of the proposed system is known as the *interface* module. Various components of this module are also described in this Section.

### 3.1 Traffic Model and Attack Model

In the proposed traffic model *packets* from the network refers to small independent queries to the server (e.g., a small HTTP query or an NTP question-answer). For simplicity, it is assumed that every query causes the same workload on the server. Since the query packets cause workload on the server, after a certain time the server cannot handle incoming traffic any further due to memory and processing overloads.

Let us suppose that the attacker uses $A$ number of hosts during the attack. When $A = 1$, the attack originates from a single source, and when $A > 1$, it corresponds to a distributed attack. There are one or more human attackers behind the attacking sources. These attacking sources are machines on the Internet controlled (taken over) by the attacker. It is assumed that the attacking machines use real addresses, and they can establish normal two-way communication with the server, like a host of any legal client. The human attacker hides behind the attacking machines in the network, which means that after carrying out the attack and after removal of all compromising traces of attack on the occupied machines, there is no way to find a trace leading to him/her.

Two types of sources are distinguished: *legal* sources and *attacking* sources. There are $N(t)$ legal sources and $A(t)$ attacking sources in time slot $t$. In the proposed model, the attacker can reach his/her goal only if the level of attacking traffic is high enough as compared to the level under normal operation. It is assumed that the attacker can control the extra traffic by changing the number of attacking machines and the traffic generated by these machines. It is also assumed that the attacker is powerful and can distribute the total attacking traffic among attacking machines at his/her choice. The reason for using several attacking machines is to make it more difficult for the server to identify and foil them. However, when the attacker uses more machines, it becomes more difficult for him/her to hide the attack. Therefore, the attacker needs to keep the number of attacking hosts at a small value, i.e., $A(t)$ should not be too large.

### 3.2 The Interface Module

A DDoS *interface* module is attached to the server at the network side. The interface module may be a software component of the server, a special-purpose hardware in the server host, or an autonomous hardware component attached to the server.

The incoming traffic enters a FIFO buffer. For the purpose of modeling and analysis a *discrete time model* is assumed. Traffic is modeled and processed over unit time slot. The server CPU processes $\mu$ storage units per time slot from the buffer. Since the buffer is fed by a random traffic, there is a non-zero probability of an event of buffer overflow. When a DDoS attack is launched, the incoming traffic quickly increases and the buffer becomes full. At this time, most of the incoming packets will be dropped and the attacker becomes successful in degrading the quality of service of the server. However, the server host will not be completely disabled at this point of time. The goal of the interface module is to effectively identify and disrupt the traffic from the attacking sources so that the normal level of service may be restored.

It is assumed that there are two states of the incoming channel: the *normal state*, and the *attack state*. While in the normal state, there is no DDoS attack on the server, in the attack state, the server is under a distributed attack. Let us assume that the attack begins at time $t^*$, and at time $t^* + \delta$, the interface buffer becomes full. At this time, the TCP modules running at the legal clients and the attacking hosts observe that no (or very few) acknowledgements are being sent back by the server. In order to defend against the DDoS attack, the first task is to detect the point of commencement the attack by making a reliable estimation of the time $t^*$.

Once the time of commencement of the attack is estimated, the next task is to identify the sources of the attack, and to disrupt the traffic arriving from these sources to the server. In the proposed scheme, this identification has been done based on the statistical properties of the traffic flow. The interface module at the server identifies all active traffic sources, measures the traffic generated by these sources, and classifies them into different sets. In order to get reliable measurements of the traffic level, these measurements are carried out during time slots between $t^*$ and $t^* + \delta$. Consequently, the effectiveness of the mechanism is heavily dependent on the time duration $\delta$. During the time $\delta$, the traffic flow between the sources and the server is not affected, i.e., the interface module in the server does not disrupt traffic from the attack sources. It is obviously desirable to have a large value for the time duration $\delta$ so that more time is available for traffic measurement. A large value of $\delta$ can be effectively achieved by using a very large buffer size. It is assumed that the total buffer size $(L)$ of the server consists of two parts. The first part $(L_1)$ is designed to serve the normal state of the server. The size of $L_1$ is chosen according to the service rate of the server and the normal probability of packet loss due to the event of a buffer overflow. The size of $L_2$ corresponds to the excess size of the buffer introduced with the purpose of gaining enough time for traffic measurements during the start-up phase of the attack for identification of the attack sources.

It is assumed that the attack begins at time $t^*$, i.e., all the attacking sources start sending packets at this time onwards. It is also assumed that the network was in normal state at any time $t < t^*$. Let $\hat{t}$ denote the expected value of $t^*$. For the sake of simplicity, it is assumed that the set of active sources is constant during the period of the attack. Let $T_n(t)$ be the aggregate network traffic from the legal sources (i.e., the normal network traffic), and $T_a(t)$ be the aggregate of the attacking traffic. Let the mean (per time slot) values of the normal and the attack traffic are $\lambda_n$ and $\lambda_a$ respectively.

$$E(T_n(t)) = \lambda_n \quad E(T_a(t)) = \lambda_a \tag{1}$$

Similarly, let the corresponding standard deviations be denoted by $\sigma_n$ and $\sigma_a$. Let $Q$ denote the *apriori* unknown ratio between $\lambda_n$ and $\lambda_a$, i.e. $Q = \lambda_a / \lambda_n$. As the time of commencement of attack $(t^*)$ is earlier than the time of its detection ($\hat{t}$), some precious time is wasted that cannot be used for traffic measurements. To minimize, this loss, the aggregate traffic level is estimated continuously by using a *sliding window* technique. The interface module in the server handles two *sliding time windows*. The longer window has a capacity of $w_l$ slots, and the shorter one has a capacity of $w_s$ slots. In this way, both an *extended-time average* level $\overline{\lambda}$ *(t)* and a *short-time average* level $\hat{\lambda}$ *(t)* of the incoming aggregate traffic per slot at time slot *t* are estimated.

### 3.3  Algorithms of the Interface Module

The interface module in the server executes two algorithms in order to identify the DDoS attack and the attacking sources. The algorithms are: (i) algorithm for detection of an attack, (ii) algorithm for identification of the attack sources and disruption of traffic arriving from the attack sources. In the following the algorithms are described.

#### 3.3.1  Algorithm for Attack Detection
In order to ensure high availability of the server, an early detection of an attack is of prime importance. As discussed in Section 3.2, the beginning of an attack is assumed to take place at time $\hat{t}$. An approximate determination of $\hat{t}$ can be done in any of the following two ways: (i) $\hat{t}$ is the point of time when the buffer $L_l$ becomes full. (ii) $\hat{t}$ is the point of time when the following inequality holds:

$$\hat{\lambda}(\hat{t}) > (1+r)\overline{\lambda}(\hat{t}) \tag{2}$$

In the inequality (2), $r > 0$ is a design parameter. It represents the maximum value of the fraction by which the short-term average of traffic level may exceed the long-term average without causing any alarm for attack on the server. In Section 4, the comparative analysis of the effectiveness of the two approaches in detecting a distributed attack is presented with simulation results. However, for a more accurate and reliable identification of an attack, a statistical approach based of hypothesis testing is also proposed. In this approach, a large sample of packet arrival pattern on the server is taken for a long duration. The *packet arrival rate* (PAR) at each sample duration is periodically measured and the sample mean ($\overline{X}$) and the sample standard deviation ($\hat{S}$) of the PAR are computed using (3) and (4).

$$\overline{X} = \frac{\sum_{i=1}^{N} X_i}{N} \tag{3}$$

$$\hat{S} = \sqrt{\frac{\sum_{i=1}^{N}(X_i - \overline{X})^2}{N-1}} \tag{4}$$

After the computation of $\overline{X}$ and $\hat{S}$, one-sample *Kolmogorov-Smirnov* (*K-S*) test is applied to test if the samples come from a population with a normal distribution. It is found that *P*- values for all *K-S* tests are greater than α = .05. Therefore, it is concluded that the PAR follows a normal distribution. In other words, $\overline{X}$ is normally distributed with an unknown mean, say, μ. The standard value of $\overline{X}$ is given by (5):

$$Z = \frac{\overline{X} - \mu}{\hat{S}/\sqrt{N}} \tag{5}$$

In (5) *Z* is a standard normal variable and satisfies (6):

$$P\{-Z_{\alpha/2} \leq \frac{\overline{X} - \mu}{\hat{S}/\sqrt{N}} \leq Z_{\alpha/2}\} = 1 - \alpha \tag{6}$$

In equation (6) α is the level of confidence which satisfies $0 \leq \alpha \leq 1$. From (6) it is clear that there is a probability of 1- α of selecting a sample for which the confidence interval will contain true value of μ. $Z_{\alpha/2}$ is the upper 100 α/2 percentage point of the standard normal distribution. The 100(1- α)% conf. interval of μ is given by (7):

$$\overline{X} - Z_{\alpha/2} \frac{\hat{S}}{\sqrt{N}} \leq \mu \leq \overline{X} + Z_{\alpha/2} \frac{\hat{S}}{\sqrt{N}} \tag{7}$$

The confidence interval in equation (7) gives both a lower and an upper confidence boundary for μ. To detect an attack scenario, a threshold value called *maximum packet arrival rate* (MPAR) is defined which distinguishes the normal PAR and the high PAR in an attack. In order to find MPAR, the upper confidence bounds for μ in (7) are obtained by setting the lower conf. bound to -∞ and replacing $Z_{\alpha/2}$ by $Z_\alpha$. A 100(1-α)% upper confidence bound for μ is obtained from (8). The value of α in (8) is 0.025.

$$\mu \leq T_x = \overline{X} + Z_\alpha \frac{\hat{S}}{\sqrt{N}} \tag{8}$$

Let $\mu_1$ and $\mu_2$ denote the population means of two traffic flows. The *t*-test is applied to determine the significance of the difference between the two means, i.e. ($\mu_1 - \mu_2$). Let the difference between the two means be ($\overline{X}_1 - \overline{X}_2$), and the standard deviation of the sampling distribution of differences is $\sqrt{(\frac{S_1^2}{N_1} + \frac{S_2^2}{N_2})}$. The *t*-statistic is computed in (9).

$$t = \frac{\overline{X}_1 - \overline{X}_2}{\sqrt{(\frac{S_1^2}{N_1} + \frac{S_2^2}{N_2})}} \tag{9}$$

Since the two groups may contain different sample sizes, a weighted variance estimate *t*-test is used. The weighted variance is computed in equation (10):

$$\hat{S}^2 = \frac{(N_1-1)S_1^2 + (N_2-1)S_2^2}{N_1+N_2-2} \tag{10}$$

The resultant *t*-statistic is computed in equation (11):

$$t = \frac{\overline{X}_1 - \overline{X}_2}{\sqrt{\frac{\hat{S}_1^2}{N_1} + \frac{\hat{S}_2^2}{N_2}}} \tag{11}$$

To detect attack traffic, the following hypotheses are tested. The null hypothesis $H_0$: $\mu_1 = \mu_2$ is tested against the alternative hypothesis $H_1$: $\mu_1 \neq \mu_2$. Levene's test is used to assess $H_0$. If the resulting *P*-values of Levene's test is less than a critical value (0.05 in this case), $H_0$ is rejected and it is concluded that there is a difference in the variances of the populations. This indicates that the current traffic flow is an attack traffic. As will be evident in Section 4.2, this accurate statistical algorithm has 100% detection accuracy in all simulation runs conducted on the system.

### 3.3.2 Algorithm for Identification of Attack Sources

It is essential to disrupt the traffic emanating from the attack sources at the interface module of the server after an attack is detected. For this purpose, the interface module must be able to distinguish between the traffic from the attack sources and the normal traffic from legitimate client hosts. It is assumed that the interface module can measure the traffic characteristics of all the active sources at each time instance by recognizing their network addresses. Starting at time $\hat{t}$, the traffic level corresponding to every source is measured. If an attack was correctly identified, i.e. $t^* < \hat{t} < t^* + \delta$, traffic measurement and analysis can be made over the period ($t^* + \delta - \hat{t}$). Let the aggregate level of traffic be $\hat{\lambda}_r (t^* + \delta)$, and the traffic for the source $i$ be $\hat{\lambda}(i) (t^* + \delta)$. As the exact traffic from the legal sources during the attack cannot be determined, the expression $\overline{\lambda}(\hat{t} - c)$, $(c > 0)$, is used as an estimate of mean aggregate traffic level of the legal sources in time interval $[t^*, t^* + \delta]$, and an estimate for the mean aggregate traffic level of the attacking sources ($\overline{\lambda}_a$) is derived as in equation (12):

$$\overline{\lambda}_a = \hat{\lambda}_r(t^*+\delta) - \overline{\lambda}(\hat{t}-c) \tag{12}$$

The set *Z* of active sources is decomposed into two mutually disjoint sets $Z_n$ and $Z_a$, where the former is the set of *legal sources* and the latter is the set of *attacking sources*. The sets *Z*, $Z_n$ and $Z_a$ will satisfy equation (13):

$$Z = Z_n \cup Z_a \quad Z_n \cap Z_a = \phi \tag{13}$$

The identification algorithm produces as output a set $Z_a^*$, which is a subset of the set $Z$ and very closely resembles the set $Z_a$. The closer the sets $Z_a$ and $Z_a^*$ are, the more accurate is the detection of the sources of attacks. The identification of the attacking sources is made by the following two ways:

(i) In this approach, the maximal subset of $Z_a^* = \{i_1, i_2, \ldots i_L\}$ of $Z$ is computed that corresponds to sources with the highest measured traffic levels so that the inequality (14) is satisfied. The set $Z_a^*$ contains the attack sources.

$$\sum_{j=1}^{v} \hat{\lambda}^{(ij)}(t^*+\delta) \leq \hat{\lambda}_a \qquad (14)$$

The basis principle for this method is that the attacker always tries to hide himself/herself, and therefore limits the number of attacking sources ($A(t)$). At the same time, to make the attack effective, the attacker intends to send a high volume of attack traffic to the server. Thus, there is a trade-off with the volume of the attack and the number of attack sources.

(ii) In this method, the sources from the set of traffic sources $Z$ which are active during the interval ($\hat{t}$ - c), c > 0, are omitted and (14) is used to identify the attack sources.

Once the attacking sources are correctly identified, the disruption of the traffic emanating from the attack sources is done. For this purpose, all the incoming packets with source addresses belonging to set $Z_a^*$ are disarded.

## 4   Simulation and Results

The simulation program is written in C and the program is run on a workstation with Red Hat Linux version 9 operating system. A MySQL database is used for storing data related to traffic. The time interval is set at $10^{-6}$ seconds. The simulation is done with first 100 seconds as the normal traffic. The attack simulation is started at the $100^{th}$ second and is allowed to continue till the $200^{th}$ second. The simulation is ended with another 100 seconds of normal traffic to test efficacy of the recovery function of the system. The traffic arrivals are modelled as Poisson process. The packets are stored in a buffer and are passed on to the CPU for further processing by the interface module. The queue type is assumed to be M/M/1. The inter-arrival time and service time are negative exponential distributions. Following cases are considered:

*Case 1*: For a small corporate server, the number of legal clients is low, say $N(t) =$ 5. Assuming that the capacity of the server is high, the average load on the server will be less. Therefore, the number of attacking hosts should be high, say $A(t) = 40$. Hence, in this scenario, for an effective attack we must have $N(t) << A(t)$.

*Case 2*: For a server of medium size, it may be assumed that $N(t) = 50$ and a successful attacker can launch his/her attack from a fewer number of hosts. Thus it may be assumed that $A(t) = 50$ in this case. As the number of legal clients and the number of attacking sources are of comparable size, it is easier for the attacker to hide his/her attack in this case. Therefore, in this situation,   $N(t) \approx A(t)$.

*Case 3*: For a global portal server, there can be a very large number of legal clients, say $N(t)$ = 10000. In this situation, it is not possible for that attacker to easily estimate the required number of attacking hosts. In this case, it is assumed that the attacker chooses a reasonably high value of $A(t)$, say $A(t)$ = 5000, and opts for a very high attacking rate: $\lambda_a = \lambda_n *10$. Therefore, in this case: $N(t) > A(t)$.

**Table 1.** Simulation parameters for Simulation I

| Parameter | Value |
|---|---|
| Number of legal clients ($N(t)$) | 10000 |
| Number of attacking hosts ($A(t)$) | 5000 |
| Mean normal traffic rate ($\lambda_n$) | 0.1 |
| Mean attack traffic rate ($\lambda_a$) | 0.4 |
| Service rate ($\mu$) (packets/sec) | 1500 |

The simulation parameters are listed in Table 1. With 10000 legal clients and $\lambda_n$ = 0.1, the capacity of the server should be at least 1000. However, the attack is successful only when the service rate ($\mu$) is less than 3000 ($\lambda_a*A(t) + \lambda_n*N(t)$). The value of $\mu$ is, therefore, taken as 1500. The buffer size for normal situation is taken as 40 packets i.e., $L_1$ = 40 (packets). For choosing the size of $L_2$, it is observed that the normal traffic rate is 1000 packets/sec. Thus a safe value of $L_2$ = 3000 (packets) is taken. The values of the parameters of the detection algorithm are given in Table 2. The available time for traffic analysis depends on the value of $\delta$. In the simulation work, a constant value ($\hat{\delta} \leq \delta$) for this parameter is used for traffic analysis. It is assumed that the total traffic (normal and attack) is known and its value is $T_n + T_a$ = 3000. As the service rate ($\mu$) is 1500, one can expect the buffer $L_1$ to be full after 40/(3000-1500) ≈ 0.3 seconds. The whole buffer ($L= L_1 + L_2$) will be full in 30040/(3000-1500) ≈ 200 seconds. Therefore, a safe estimation of $\hat{\delta}$ = 10 is made. In real world situation, $\delta$ should be estimated over a period of time. For simplicity, the value of $\hat{\delta}$ is set equal to $w_s$. The algorithm presented in Section 3.3.2 is used for identification of the attacker.

**Table 2.** Parameters of the attack detection algorithm

| Parameter | Value |
|---|---|
| Sliding window size ($w_s$) | 10 sec |
| Tolerance for traffic jump ($r$) | 0.6 |
| Time frame for last correct value of $\lambda$ | 45 sec |

### 4.1 Simulation I

Table 3 shows the results of the simulation with different values of the window size ($w_s$). It is clear that a larger window size and hence a large $\delta$ gives a more accurate identification of attacks. However, with a larger window size the system is more likely to enter into a situation of buffer overflow. After the buffer overflow, the detection algorithm will produce very inaccurate and unreliable results. Therefore it is not worthwhile to increase the window size beyond a limit. On the other hand, when the time window is too short, the algorithm can detect only a very small proportion of the attacking hosts. In summary, the results In Table 3 show that the mechanism can detect an attack with a window size of 10 seconds.

**Table 3.** Results of Simulation I

| Observed metrics | $\hat{\delta}$ ($\hat{\delta} = w_s$) | | | | |
|---|---|---|---|---|---|
| | 5 | **10** | 20 | 30 | 40 |
| Correctly identified attackers | 2982 | **3784** | 4529 | 4784 | 4892 |
| Filtered legal clients | 1 | **557** | 260 | 132 | 59 |
| Dropped packets | 0 | **0** | 0 | 14251 | 28765 |
| Max. buffer level and corresponding time frame | 29717 (200 s) | **14941 (110s)** | 29732 (119s) | 30040 (120s) | 30040 (120s) |
| Time to restore (after $t^*$) | 149 | **104** | 73 | 71 | 81 |

### 4.2 Simulation II

In this case, a smaller system is simulated with parameters are listed in Table 4. The buffers $L_1$ and $L_2$ are chosen as 40 and 160 respectively. The value of $\delta$ is set equal to $w_s$, i.e. $\hat{\delta} = w_s = 10$. The remaining parameters are kept the same as in simulation I.

In simulation II, experiments are repeated on 500 different sets of input data to have an insight into the statistical properties of the system under normal and attack situations. With different data sets, it is observed that the approximate algorithm (ii) in Section 3.3.1 was faster in detecting the attack in 454 cases. In 42 cases, the attack was correctly identified by both algorithms (i) and (ii) in Section 3.3.1. The accurate detection algorithm presented in Section 3.3.1 could detect all the 50 attack sources in all the 500 simulation runs. Table 5 summarizes the simulation results.

**Table 4.** Parameters for Simulation II

| Parameter | Value |
|---|---|
| Number of legal clients ($N(t)$) | 50 |
| Number of attacking hosts ($A(t)$) | 50 |
| Mean normal traffic rate ($\lambda_n$) | 0.1 |
| Mean attack traffic rate ($\lambda_a$) | 0.2 |
| Service rate ($\mu$) (packets/sec) | 8 |

**Table 5.** Results of Simulation II

| Observed metrics | Observed values | | |
|---|---|---|---|
| | Min | Avg | Conf. Int. (95%) |
| Traffic restoration time (after $t^*$) | 49 | 114.732 | 1.942 |
| Packets dropped | 0 | 0.695 | 0.321 |
| Normal user filtered (type II error) | 1 | 7.115 | 0.231 |
| Number of attackers filtered | 21 | 32.413 | 0.235 |
| Attack detection time (after $t^*$) | 0 | 2.95 | 0.09 |

## 5 Conclusion

In this paper, a mechanism is presented for detection and prevention of DDoS attacks on a server. Different algorithms are presented for attack detection based on statistical theory of hypothesis testing. While the proposed mechanism does not affect the traffic from legitimate clients, it effectively blocks traffic from the attack sources. The simulation results demonstrate the effectiveness of the proposed mechanism.